\documentclass[superscriptaddress,twocolumn,showpacs,preprintnumbers,amsmath,amssymb,prl,aps]{revtex4-1}
\usepackage{verbatim,psfrag,graphicx,times,epsfig,color,bm,dcolumn,hyperref}
\usepackage[all]{xy}

\begin{document}

\title{Polarized currents and spatial separation of Kondo state: NRG study of spin-orbital effect in a double QD}
\author{E. Vernek}
\affiliation{Instituto de F\'{\i}sica, Universidade Federal de Uberl\^andia, Uberl\^andia, MG 38400-902, Brazil.}
\affiliation{Instituto de F\'{i}sica de S\~ao Carlos, Universidade de S\~ao Paulo, S\~ao Carlos, S\~ao Paulo 13560-970, Brazil}
\author{C.~A. B\"usser}
\affiliation{Department of Physics and Arnold Sommerfeld Center for Theoretical Physics, 
Ludwig-Maximilians-Universit\"at M\"unchen, D-80333 M\"unchen, Germany.}
\author{E.~V. Anda}
\affiliation{Departamento de F\'{\i}sica, Pontif\'{\i}cia Universidade Cat\'olica, Rio de Janeiro-RJ, Brazil.}
\author{A.~E. Feiguin}
\affiliation{Department of Physics, Northeastern University, Boston, Massachusetts 02115, USA.}
\author{G.~B. Martins}
\email[Corresponding author: ]{martins@oakland.edu}
\affiliation{Department of Physics, Oakland University, Rochester, MI 48309, USA.}
\affiliation{Materials Science and Technology Division, Oak Ridge National Laboratory, Oak Ridge, Tennessee 37831, USA}
\begin{abstract}
A double quantum dot device, connected to two channels that only see each other through interdot Coulomb repulsion, is analyzed 
using the numerical renormalization group technique. By using a two-impurity Anderson model, and parameter values 
obtained from experiment [S. Amasha {\it et al.}, Phys. Rev. Lett. {\bf 110}, 046604 (2013)], it is shown that, by applying 
a moderate magnetic field, and adjusting the gate potential of each quantum dot, opposing spin polarizations are created in each channel. 
Furthermore, through a well defined change in the gate potentials, the polarizations can be reversed. This polarization effect 
is clearly associated to a spin-orbital Kondo state having a Kondo peak that originates from spatially separated parts of the device. 
This fact opens the exciting possibility of experimentally probing the internal structure of an SU(2) Kondo state. 
\end{abstract}
\maketitle

{\it Introduction.}\textemdash 
Quantum dots (QDs) \cite{Goldhaber-Gordon1998} 
 provide unprecedented experimental control over all parameters leading to the many-body Kondo state \cite{Hewson1997}. 
The last decade has seen a remarkable sequence of new developments, mainly in lateral semiconducting QDs \cite{hanson2007}: 
observation of an SU(4) regime with entanglement of spin and orbital degrees 
of freedom \cite{Jarillo-Herrero2005,Keller2013}, a non Fermi liquid ground state \cite{Potok2007}, and, 
more recently, the careful analysis of a double QD (DQD) Kondo effect with both spin and orbital degrees of freedom \cite{Amasha2013}. 
In the work presented here, based on this latter DQD device, we investigate an `exotic' Kondo state, where {the electron spin in the `traditional'
Kondo effect \cite{Goldhaber-Gordon1998} is replaced by a so-called pseudospin. Still, the symmetry of the system's Hamiltonian is SU(2), and hence the effective
low energy model is an SU(2) Kondo model.
Indeed, an SU(2) Kondo state requires a system where: {\it (i)} the relevant degrees of freedom have SU(2) symmetry, 
{\it (ii)} they can form a charge reservoir consisting of itinerant states, {\it (iii)} 
have at least one localized state that has an overlap with the itinerant states, 
{\it (iv)} double occupancy of a localized state should cost repulsion energy. 
In the original observation of the Kondo effect \cite{Hewson1997} in solids, the repulsion energy was intraorbital
Coulombic interaction, while in the recently man-made QD devices \cite{Goldhaber-Gordon1998}, it is mesoscopic charging energy.  Nonetheless,
the low energy physics in both systems is captured by a Kondo model in which the relevant degree of freedom is the electronic spin.
A superexchange interaction between localized and itinerant states
results in the fluctuating localized spin being screened by itinerant spins,
forming a many-body spin-singlet state. In both cases,
the Kondo Hamiltonian is written in terms of the electron spin operator. 

\begin{figure}
  \begin{center}
    \includegraphics[width=3.2in]{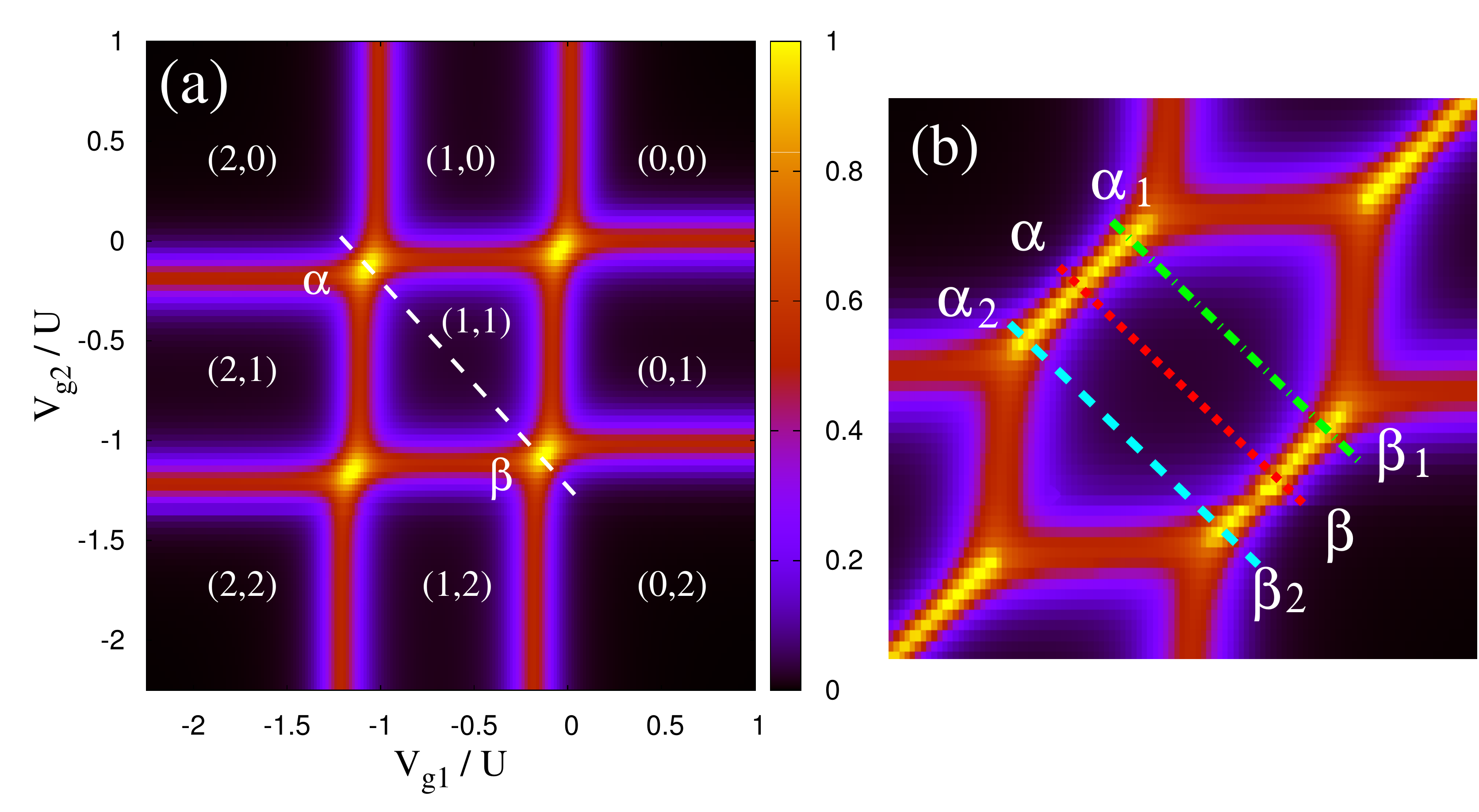}
  \end{center}
\caption{(Color online) 
(a) Contour plot of the charge stability diagram showing the conductance 
(in units of $2e^2/h$) in the $V_{g1}-V_{g2}$ plane. Charge occupancies are indicated as $\left(n_1,n_2\right)$.
Parameter values are $U^{\prime}=0.1$, $H=0.013$, and $\Gamma=0.027$. The conductance was obtained by using DMRG to calculate 
the charge occupancy $n_1$ and $n_2$ for each QD, followed by the application of the 
FSR \cite{Busser2012}. The dashed (white) line is 
defined by $V_{g2}=-V_{g1}-(U+2U^{\prime})$. Most of the results 
shown in this work are for the Kondo regime at point $\beta$. 
(b) Zoom in on the central region of panel (a) showing the three points ($\beta_1$, $\beta$, and $\beta_2$) 
where the Kondo state is analyzed [details are shown in the 
inset to Fig.~\ref{figure3}(a)].
}
\label{figure1}
\end{figure}

In an `exotic' SU(2) Kondo state, although the effective low energy model 
is as described above, the localized and itinerant SU(2) degrees of freedom 
may be more elaborate. Indeed, the operator describing the SU(2) degree of freedom, say $\tau^z$, may be a function 
of operators whose eigenvalues probe spatially distinct parts of the system. In the following, 
we will argue that this indicates that the many-body Kondo state possesses an `internal structure'.
The `structure' of the operator $\tau^z$, i.e. its explicit dependence 
on experimentally observable quantities, may be tailored in a way that results in 
novel effects (spin polarized conductance, for instance). More importantly, this structure 
may enable experimentalists to manipulate the Kondo state, perhaps resulting in new device 
functionality. In this work, we describe such an `exotic' Kondo state, where the pseudospin SU(2) degree of freedom 
$\tau^z\left(n_1,n_2\right)$ is a {\it function} of the charge occupancy of two 
capacitively coupled QDs connected to two spatially separated charge reservoirs. 
Using the Numerical Renormalization Group (NRG) \cite{Krishna-murthy1980} and the Density Matrix Renormalization Group (DMRG) 
methods \cite{White1992}, we show that this internal structure translates into spin polarized currents with 
opposite polarizations in each channel [see Fig.~\ref{figure2}(b)] and into a spatial separation of the local density of states 
(LDOS) at the Fermi energy (i.e. spatial separation of the Kondo resonance). 
Finally, our calculations, made for realistic parameter values \cite{Amasha2013}, show that 
the spin polarization, and therefore the above mentioned internal structure, can be observed experimentally. For that, 
we suggest the integration of the DQD here analyzed into a ballistic spin resonator device 
capable of measuring spin currents \cite{Frolov2009,Frolov2009a}. This measurement requires the creation of a current 
through the device, and therefore the application of a bias to one of the QDs. In the inset to Fig.~\ref{figure4}(b) 
we show that the Kondo effect analyzed here is robust against asymmetry in the couplings $\Gamma_i$, for different channels 
$i=1,2$. This indicates that the Kondo effect, and the spin polarization, should be robust against a small bias. 

\begin{figure}
  \begin{center}
    \includegraphics[width=3.0in]{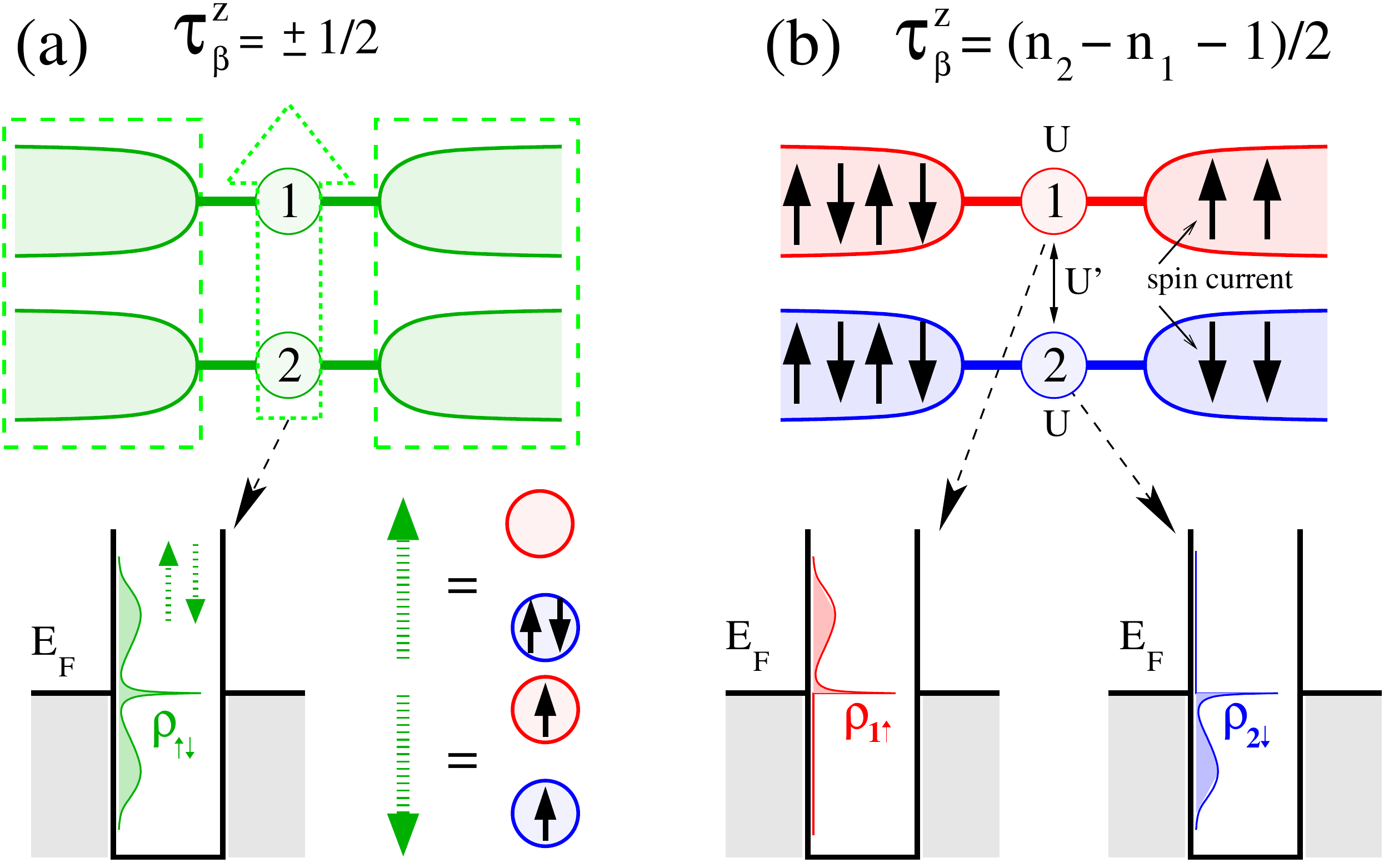}
  \end{center}
\caption{(Color online) (a) Top: schematic representation of the pseudospin $\tau_{\beta}^z$ 
Kondo effect, where spin and orbital degrees of freedom for {\it both} QDs participate in the Kondo state. 
This is depicted by the (green) dotted arrow superposed to both QDs, and the (green) dashed boxes containing both leads. 
Bottom left: pseudospin LDOS, showing the SU(2) Kondo peak pinned to the Fermi energy. 
The dashed (green) arrows picture the pseudospin fluctuations, a hallmark of the Kondo effect. 
Bottom right: legend relating  $\tau_{\beta}^z=\pm 1/2$ to $\left(n_1,n_2\right)$. 
(b) Top: analysis of the Kondo state into its {\it constituent parts}. Noting 
that solid arrows represent {\it electron spins}, the currents with opposing spin polarization in each channel, 
generated by the Kondo effect, are depicted in the right leads. Note that the pseudospin $\tau_{\beta}^z=\left(n_2-n_1-1\right)/2$ 
is related to the DQD orbital/spin degrees of freedom. Bottom: the surprising decomposition of the Kondo peak when the experimentally 
accessible LDOS is plotted: the spin up LDOS in QD 1 has the same participation in the Kondo peak as the spin down LDOS of QD 2 and they 
originate in spatially distinct parts of the DQD system. 
}
\label{figure2}
\end{figure}

{\it Setup and two-impurity Anderson Hamiltonian.}\textemdash We consider capacitively coupled parallel QDs \cite{Okazaki2011,Amasha2013}
connected to metallic leads that are correlated to each other only through the interdot coupling $U^{\prime}$ [Fig.~\ref{figure2}(b)].
Through an even-odd transformation \cite{Busser2011}, two leads decouple from the DQD and the electron 
reservoirs are reduced to two non-interacting semi-infinite chains [see Eq.~(3)].
Thus the two-impurity Anderson Hamiltonian (TIAM) $H_{\rm tot} = H_{\rm dqd} + H_{\rm band} + H_{\rm hyb}$ modeling our system is given by 
\begin{eqnarray}
H_{\rm dqd}&=&\sum_{\lambda;\sigma}
\left[ {U \over 2} n_{\lambda \sigma} n_{\lambda \bar{\sigma}} + V_{g\lambda} n_{\lambda \sigma}\right] + U^{\prime} \sum_{\sigma \sigma^{\prime}} n_{1 \sigma} n_{2 \sigma'}\\
  \label{eq1}
 H_{\rm band} &=&  -t \sum_{\lambda}
 \sum_{i=1;\sigma}^\infty
  (c_{\lambda i\sigma}^{\dagger} c_{\lambda i+1\sigma} +\mbox{H.c.}) \\
  \label{eq2}
H_{\rm hyb} &=& -\sum_{\lambda;\sigma} t_{\lambda}
\left[ d_{\lambda \sigma}^{\dagger} c_{\lambda1\sigma} + \mbox{H.c.} \right].
  \label{eq3}
\end{eqnarray}
The operator $d_{\lambda \sigma}^\dagger$ ($d_{\lambda \sigma}$) creates (annihilates)
an electron in QD $\lambda=1,2$ with spin $\sigma=\pm$, while operator $c_{\lambda i \sigma}^\dagger$
($c_{\lambda i+1 \sigma}$) does the same at site $i$ ($i+1$) in a non-interacting semi-infinite chain $\lambda=1,2$;
$n_{\lambda \sigma}= d_{\lambda \sigma}^\dagger d_{\lambda \sigma}$ is the
charge per spin at each QD, both QDs having the same charging energy $U$ \cite{Amasha2013}, 
while charges in different QDs interact through a capacitive coupling $U^{\prime}$. 
We also include the effect of a magnetic field through $-\sum_{\lambda}gHS_{\lambda}^z$ (with $g=2.0$) acting just on the QDs \cite{Recher2000},
and coupling only to the spin degree of freedom $S_{\lambda}^z=\pm 1/2$ of each QD \cite{note-orb}.
And, for simplicity, we take $t_1=t_2=t^{\prime}$ \cite{Amasha2013}.  
This model has been extensively studied in previous works,
and it is well known that for $U^{\prime}/U=1.0$ and zero-field it has an SU(4) Kondo fixed point \cite{Galpin2006,Busser2007}, 
experimentally observed in Refs.~\cite{Jarillo-Herrero2005,Makarovski2007}. 
It is important to note that, contrary to Ref.~\cite{Feinberg2004} (where the Kondo state here analyzed was first discussed), 
the {\it only} interaction between electrons in different channels $\lambda=1,2$ is the interdot coupling $U^{\prime}$.
Moreover, each QD has independently controllable gate potentials $V_{g1}$ and $V_{g2}$, and we will concentrate 
on the experimentally relevant regime $U^{\prime}/U = 0.1$ \cite{Okazaki2011,Amasha2013}, where $U$ is our unit of energy.
The width of the one-body resonance for each QD is taken as $\Gamma= 0.027$, where $\Gamma= \pi t^{\prime 2} \rho_{0}(E_F)$,
and $\rho_{0}(E_F)$ is the density of states of the leads at the Fermi energy $E_F$. We set the half bandwidth 
$D=1.0$, and the NRG-estimated \cite{lambda} Kondo temperature $T_K=0.5K$ was obtained for these experimental parameter values \cite{kondo-temp}. 

{\it DQD charge stability diagram.}\textemdash Figure \ref{figure1}(a) shows the 
DQD charge stability diagram \cite{Wiel2002} (at finite field $H=0.013$ and $U^{\prime}=0.1$) through a conductance contour plot, as a function of the 
gate potentials $V_{g1}$ and $V_{g2}$. $(n_1,n_2)$ (where $n_{\lambda}=\sum_{\sigma}{n_{\lambda \sigma}}$) shows the charge occupation 
for each Coulomb blockade valley. The results in Fig.~\ref{figure1}(a) where obtained through a 
combination of DMRG \cite{White1992} and the Friedel Sum Rule (FSR) \cite{Busser2012}.
The separation between the two parallel bright (yellow) lines [with unit conductance, and shown 
in more detail in panel (b)] depends on $U^{\prime}$ and $H$. 
Indeed, the zero conductance region between them (where $n_1=n_2=1$) is obtained by applying a magnetic field $H=0.013$, which 
suppresses the spin Kondo effect. In addition, their lengths are proportional to $U^{\prime}$. 
The (white) dashed line in Fig.~\ref{figure1}(a), defined by $V_{g2}=-V_{g1}-(U+2U^{\prime})$, determines points $\alpha$ and $\beta$ as the 
values of $V_{g1}$ and $V_{g2}$ (over this line) for which the conductance is a maximum, coinciding with charge degeneracy points. 
Around these points one can define a pseudospin operator \cite{Busser2012,Feinberg2004} that depends solely on $n_1$ and $n_2$: $\tau_{\alpha}^z=(n_1-n_2-1)/2$ 
and $\tau_{\beta}^z=(n_2-n_1-1)/2$, so that $\tau_{\alpha}^z= 1/2$, for $(2,0)$, and $-1/2$ for $(1,1)$; 
$\tau_{\beta}^z= 1/2$ for $(0,2)$, and $-1/2$ for $(1,1)$.  
Note the important point that, out of the four possible different $\left(1,1\right)$ spin states, the application of 
$H$ creates a doublet ground state, $\left(0,\uparrow\downarrow\right)$ and $\left(\uparrow,\uparrow\right)$ at point $\beta$, 
which forms the basis for the spin-orbital Kondo state [see Fig~\ref{figure2}(a), bottom panel]. 
Indeed, a straightforward calculation at point $\beta$, to zero order in $\Gamma$, indicates that these states are degenerate. 
In addition, the states with $n_1+n_2=1$ and $3$, $|0,\uparrow\rangle$ 
and $|\uparrow,\uparrow\downarrow \rangle$, respectively, are above this ground state doublet by exactly $U^{\prime}/2$ 
(again, to zero order in $\Gamma$). These are the virtual states, participating in cotunneling processes, 
that lead to the quenching of the pseudospin (Kondo effect) {\it and} to the current polarization. 
Note that the $\alpha - \beta$ line bisects the charge degeneracy lines, therefore points $\alpha$ and $\beta$ have symmetrical 
LDOS around the Fermi energy (see Fig.~\ref{figure4}). But, other than that, other points over the charge 
degeneracy line also display Kondo states [points $\beta_{1,2}$ in Fig.~\ref{figure1}(b) are analyzed below]. 

{\it Kondo state internal structure.}\textemdash Before describing the NRG results, we 
wish to explain what is meant by {\it spatial separation} 
of the Kondo state: at the top of  Fig.~\ref{figure2}(a) we depict, through a dotted (green) arrow, the $\tau_{\beta}^z=+1/2$ projection of 
the pseudospin in the low-energy effective SU(2) Kondo model describing the spin-orbital Kondo effect \cite{Busser2012,Feinberg2004}. 
This arrow is superimposed to both QDs to indicate that the pseudospin is composed by a combination of degrees of freedom from both QDs. 
For the same reason, the charge reservoirs are circumscribed by (green) dashed boxes. 
At the bottom left side of panel (a) we depict the traditional LDOS of an SU(2) Kondo state at the 
particle-hole symmetric point (PHSP), with a symmetric 
Kondo peak at the Fermi energy. Panel (b) breaks down the pseudospin (see equation at the top), and the Kondo LDOS 
(diagrams at the bottom), into its constituent parts. 
As expected, $\rho_{1\uparrow} + \rho_{2\downarrow}$ for the two diagrams in the bottom is approximately equal to the LDOS in the bottom left 
diagram. What is interesting is that $\rho_{1\uparrow}$ and $\rho_{2\downarrow}$ are mirror reflections of each other around 
the Fermi energy (see Fig.~\ref{figure4} for details), and, as indicated by the dashed arrows, they originate from spatially distinct parts of the system. 
In addition, to the right of the `traditional' SU(2) LDOS, the $\tau_{\beta}^z=\pm 1/2$ pseudospin states [dashed (green) arrows] 
are described in terms of orbital and spin degrees of freedom. As mentioned above, these states are degenerate at point $\beta$.  
Finally, the spin polarization of the current in each channel is depicted in the right side leads on the top of panel (b). Note that 
the direction of the spin polarization of the current in each channel matches that of the finite LDOS in each of the quantum well diagrams. 
This neatly links the spatial separation of the Kondo peak with the spin polarization of the current in each channel. 
One could then say that this spatial separation of the Kondo peak (reflected in the different current polarizations in each channel) lays bare 
to the experimentalist what we loosely call the `internal structure' of the Kondo state. 

\begin{figure}
  \begin{center}
    \includegraphics[width=3.0in]{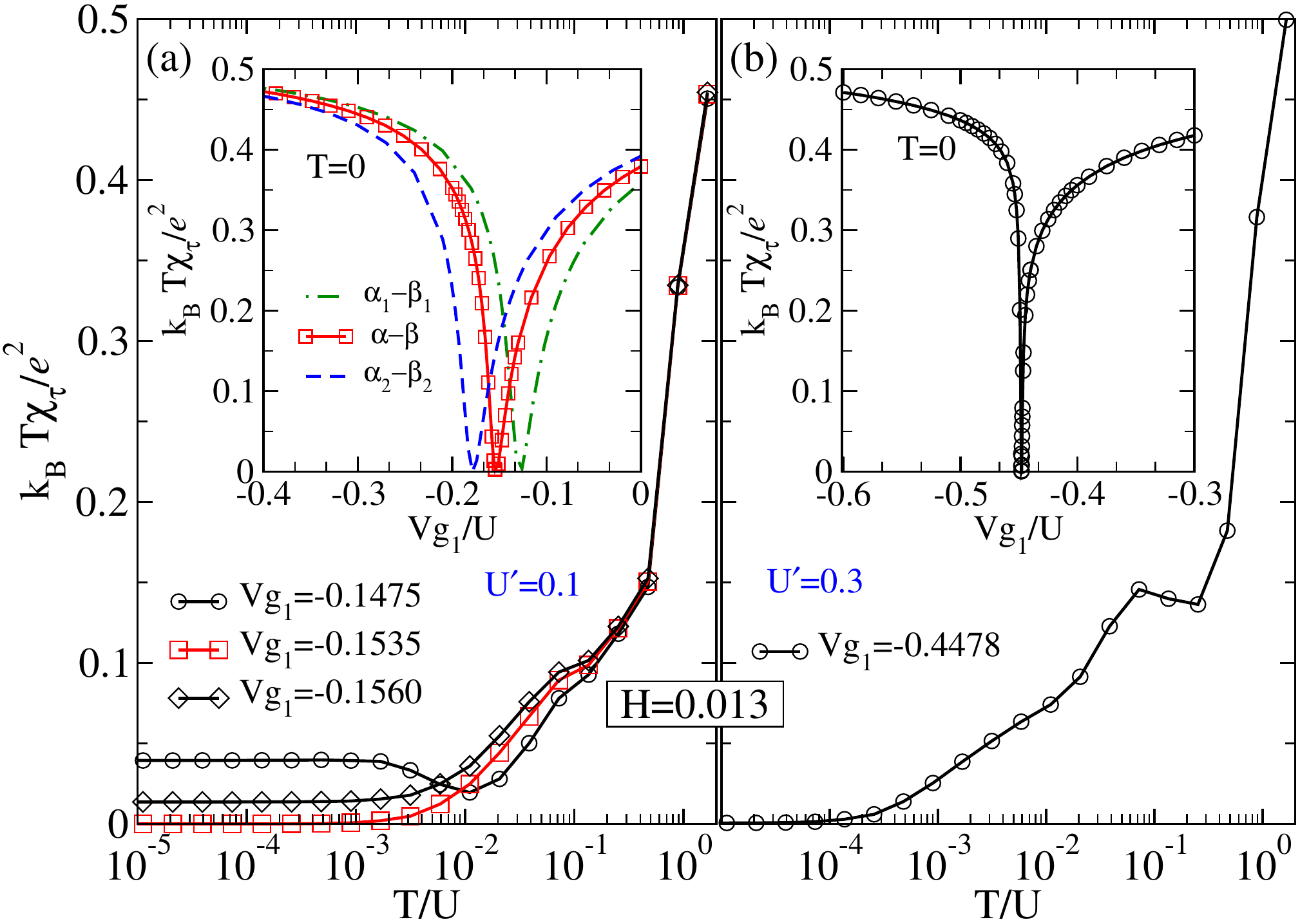}
  \end{center}
\caption{(Color online) (a) Variation with temperature (in units of $U$) 
of $\chi_{\tau^z}$, indicating the Kondo screening of $\tau_{\beta}^z=(n_2 - n_1 -1)/2$ 
at low temperatures for $V_{g1}=-0.1535$ [open (red) squares curve], 
where the conductance reaches a maximum in Fig.~\ref{figure1}(a) along the (white) dashed line (point $\beta$). 
From the open (red) squares curve the Kondo temperature (value for which $\chi_{\tau^z}=0.0701$) $T_K=0.038802$ is extracted. 
Two extra curves (for $V_{g1}$ values above and below $V_{g1}=-0.1535$) are shown in the main panel, 
$V_{g1}=-0.1475$, open circles, and $V_{g1}=-0.1560$, open diamonds, with 
finite plateaus at low temperatures for both $V_{g1}$ values. 
Inset: Variation of $\chi_{\tau^z}$ with $V_{g1}$ at zero temperature. The 
middle curve corresponds to varying $V_{g1}$ and $V_{g2}$ along the (red) 
dotted line in Fig.~\ref{figure1}(b). The other two curves correspond to the 
dot-dashed (green) and the dashed (blue) lines in the same figure (see text for details). 
(b) Similar results as in panel (a), but now for $U'=0.3$. Notice (main panel) 
the much reduced Kondo temperature $T_K=0.009370$, and the much sharper dip 
in $\chi_{\tau^z}$ as a function of $V_{g1}$ (inset), pointing to a less 
favorable experimental situation to observe the polarization, i.e., a narrower window 
in gate potential and a lower Kondo temperature. 
}
\label{figure3}
\end{figure}
{\it Thermodynamic properties.}\textemdash First, using NRG, we calculate the pseudospin susceptibility 
$\chi_{\tau^z}\left(T\right)$, equivalent to $\chi_{\tau^z}= \langle \left( \tau^z \right)^2 
\rangle-\langle \tau^z \rangle^2$ (where $\langle ... \rangle$ indicates a canonical ensemble average)
to pinpoint the Kondo effect through the screening of $\tau^z$ as $T \rightarrow 0$. 
In the main panel of Fig.~\ref{figure3}(a), we show the temperature variation of 
$\chi_{\tau^z}$ for three different $\left(V_{g1},V_{g2}\right)$ sets over the dashed 
(white) line in Fig.~\ref{figure1}(a): at point $\beta$ [$V_{g1}=-0.1535$, (red) squares in Fig.~\ref{figure3}(a)], 
and two other points away from the (yellow) charge degeneracy line [$V_{g1}=-0.1475$, (black) circles,  located inside the $\left(0,2\right)$ region, 
and $V_{g1}=-0.1560$, (black) diamonds, inside the $\left(1,1\right)$ region]. Note that for for point $\beta$, 
$\chi_{\tau^z}$ vanishes below $T \lesssim 10^{-3}$, indicating a quenching of the pseudospin $\tau^z$ due to a Kondo effect. 
The other two curves, for $V_{g1}$ values above and below the $\beta$ Kondo point, have finite-value plateaus for vanishing temperatures. 
By applying the Wilson criterion \cite{Krishna-murthy1980} for determining the Kondo temperature to the (red) squares curve in Fig.~\ref{figure3}(a), 
one obtains $T_K=0.038802$, which, for gold electrodes and the experimental $U$ value, results in $T_K \sim 0.5K$, in agreement 
with the experimental results obtained in Ref.~\cite{Amasha2013}, from where our parameters were extracted.  To illustrate the fact that the Kondo effect, 
as mentioned above, extends along the direction of the charge degeneracy (yellow) line, left and right of the 
$\beta$ point, we show, in the inset to Fig.~\ref{figure3}(a), $\chi_{\tau^z}\left(T=0\right)$ along the lines 
$\alpha_1 - \beta_1$ [dot-dashed (green) curve] and $\alpha_2 - \beta_2$ [dashed (blue) curve]. For comparison, 
$\chi_{\tau^z}\left(T=0\right)$ along the $\alpha - \beta$ line [(red) squares] is also shown. It can be seen 
that the susceptibility dips to zero in the same way along the three lines, clearly 
showing that the Kondo state is not restricted to the $\beta$ point. In addition, the polarization $P_{\lambda}\left(\beta_i\right)$ \cite{supplement} 
at the intersection of the $\alpha_i - \beta_i$ ($i=1,2$) lines with the charge degeneracy line, 
calculated with NRG, results in $P_1\left(\beta_1\right)=0.9344$, $P_2\left(\beta_1\right)=-0.6363$, 
and $P_1\left(\beta_2\right)=0.4750$, $P_2\left(\beta_2\right)=-0.9308$, indicating that the polarization 
is robust along the charge degeneracy line. 
In Fig.~\ref{figure3}(b), $\chi_{\tau^z}\left(T\right)$ results for a larger $U^{\prime}=0.3$ value 
are in qualitative agreement with a Kondo effect driven by $U^{\prime}$: as expected, the pseudospin quenching occurs 
at a lower temperature $T_K=0.009370$. 

\begin{figure}
  \begin{center}
      \includegraphics[width=2.9in]{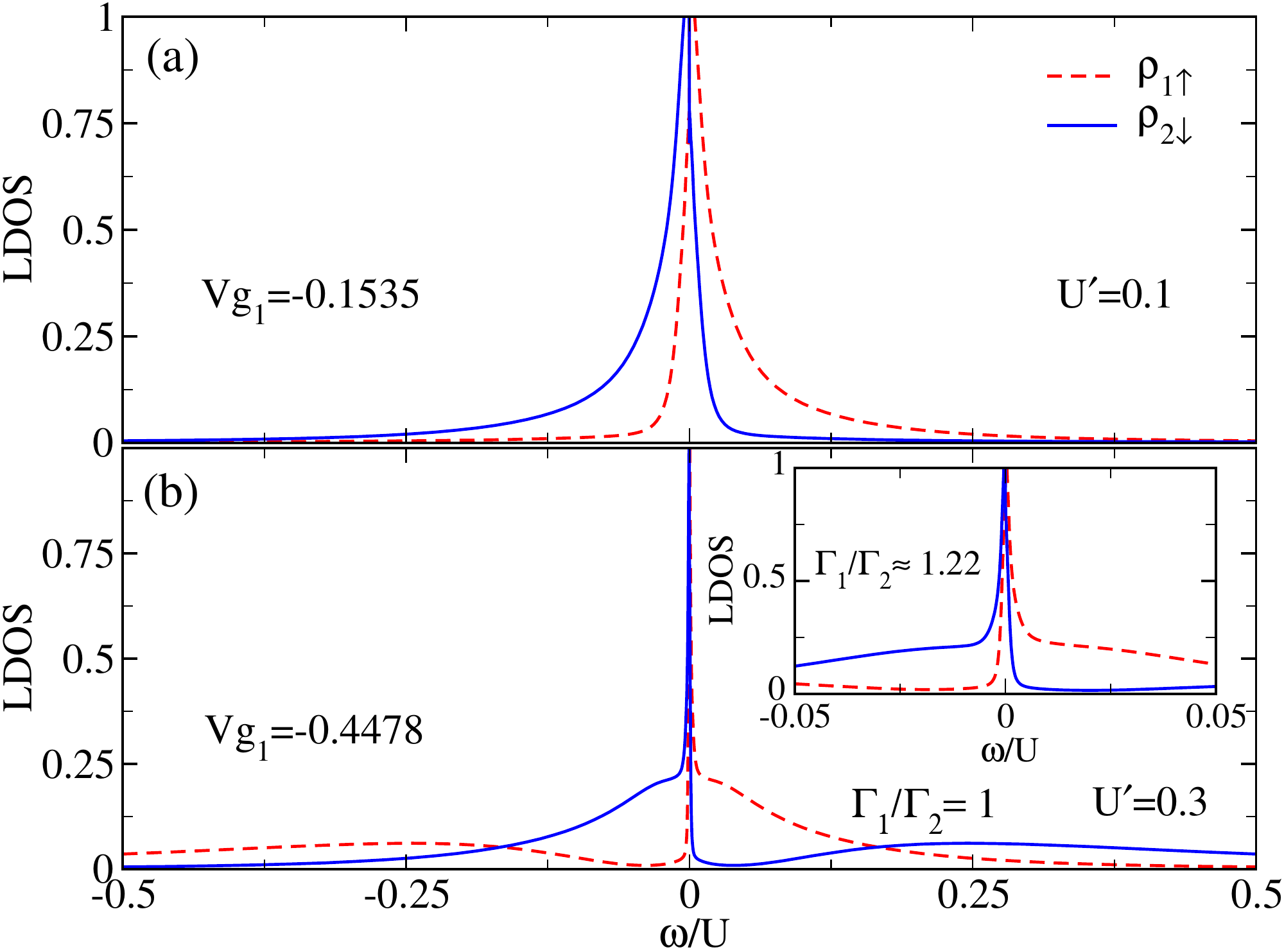}
  \end{center}
\caption{(Color online) LDOS $\rho_{\lambda \sigma}$ for both channels $\lambda$ and 
for opposite spin orientations, at point $\beta$, clearly showing a Kondo peak. 
(a) Results for $U^{\prime} =0.1$. The shape of the Kondo peaks is 
reminiscent of that for a valence fluctuation regime. (b) Same as (a), 
but for $U^{\prime}=0.3$. Note the broad peaks at $\approx \pm -U^{\prime}/2$ 
and the very particular distribution of spectral weight around the Fermi energy. 
The fact that the spectral weight is separated according to channel and spin 
orientation points to the `internal structure' of the Kondo regime. 
Inset: The particular structure of the Kondo peak (and therefore the spin polarization) 
is robust against an asymmetry in the QD/channel coupling ($\Gamma_1/\Gamma_2 \neq 1$). 
}
\label{figure4}
\end{figure}

{\it LDOS results: spatial separation of the Kondo peak.\textemdash} In Fig.~\ref{figure4} we show the main results in this work: NRG calculations 
for the LDOS in each channel, for different spin orientations, at point $\beta$. The (red) dashed curve is for channel 
1 and spin up, while the (blue) solid curve is for channel 2 and spin down [$U^{\prime}=0.1$ and $0.3$, for (a) and (b), respectively, 
with $\Gamma=0.027$ for both panels]. Note that the maximum LDOS value for each curve (in both panels) 
occurs at the Fermi energy ($\omega=0.0$). In addition, because $\beta$ is a PHSP, the curves for each spin polarization 
are mirror images of each other in relation to the Fermi energy, therefore, 
their sum produces a symmetrical curve. 
Finally, a plot of the LDOS for each `orientation' of $\tau^z$ at point $\beta$ (not shown) produces two identical curves, 
symmetrical around $\omega=0.0$, the same as the LDOS for each spin orientation in a zero-field spin SU(2) Kondo regime.
This fact hints at the main difference between the spin-orbital Kondo state studied here and a traditional {\it spin} 
SU(2) Kondo state: the latter presents to the experimentalist a localized fluctuating spin degree of freedom that is screened by 
spins carried by conduction electrons, and no easy experimental access is given to each of its projections along 
a certain direction. Indeed, the application of a magnetic field will not separate the Kondo peak into two, 
but rather split it, therefore destroying the Kondo state. On the other hand, in the spin-orbital Kondo state analyzed here, 
the fluctuating degree of freedom $\tau_{\beta}^z=(n_2-n_1-1)/2$ is, by definition, 
composed by operators from different parts of the device. This is so because the 
low energy fixed point for the TIAM at point $\beta$ is that of an SU(2) Kondo model for the {\it effective} 
$\tau_{\beta}^z$ degree of freedom. This situation is reminiscent of the SU(4) Kondo effect, 
where the same spatial separation may occur. However, the system where the cleanest observation of the SU(4) Kondo 
effect has been performed, i.e., carbon nanotubes \cite{Makarovski2007}, does not provide the experimentalist with 
easy ways to access manifestations of this spatial separation, preventing access to the Kondo state internal structure. 
In the case presented here, this internal structure is {\it given} by construction, being the opposing spin polarizations 
in each channel its most transparent manifestation. It is clear that both {\it sides} of 
the Kondo peak in Fig.~\ref{figure4} are correlated and any probing of one side, to assess one of the {\it parts} of the 
Kondo state, will affect the other. Our point is that it is the ability 
of carrying out this experiment, and analyzing 
how one `side' will affect the other, that may present a novel approach to study the Kondo state, 
and possibly have technological implications. 
Fig.~\ref{figure4}(b) shows this `asymmetry' in the LDOS for a larger $U^{\prime}=0.3$, depicting a much sharper Kondo peak, with an almost discontinuous 
drop of the LDOS at $\omega=0$, presenting a striking picture of the internal structure of the Kondo state and its 
spatial separation. 
Finally, the inset in Fig.~\ref{figure4}(b) shows LDOS results 
for a more realistic case where the coupling of the reservoirs to each QD are not equal. Indeed, allowing for $\approx 20\%$ 
difference in the couplings does not result in drastic changes in the Kondo peak. 

{\it Conclusions.}\textemdash In this work, using NRG, we have shown that a DQD device, with two channels solely 
connected by interdot Coulomb repulsion, and modeled by a TIAM with experimentally relevant parameters, 
can sustain opposite spin polarization currents along each channel. 
This current polarization effect was clearly shown to be directly related to a spin-orbital 
Kondo state (with $T_K \approx 0.5K$ for the experimental parameters) driven by the interdot repulsion $U^{\prime}$. 
Although it is necessary to apply a magnetic field to achieve the Kondo state, the moderate value needed to obtain $100\%$ polarization 
in each channel ($H \approx 1T$) \cite{gvalue} indicates that the effect should be experimentally observable \cite{Frolov2009,Frolov2009a}. 
However, the most exciting finding in this work is that this double-channel polarization can 
be clearly associated to a spatial separation of the Kondo peak, implying a spatial 
separation of the Kondo state, which can in principle lead to experimental ways of exploring 
its internal structure. The channel-specific spin polarization is the most evident fingerprint of this spatial 
separation, and it should be detectable by integrating the DQD into a Ballistic Spin Resonator device \cite{Frolov2009,Frolov2009a}. 

{\it Acknowledgements.}\textemdash We gratefully acknowledge fruitful discussions with Sami Amasha, David Goldhaber-Gordon, and Andrew Keller. 
CAB was supported by the {\it Deutsche Forschungsgemeinschaft} (DFG) through FOR 912 under grant-no. HE5242/2-2. 
EVA acknowledges the Brazilian agencies FAPERJ(CNE) and CNPq for financial support. 
GBM acknowledges financial support from NSF under Grants No. DMR-1107994 and MRI-0922811; 
and in part by the U.S. Department of Energy, Office of Basic Energy Sciences, Materials
Sciences and Engineering Division. AEF thanks NSF for support under Grant No. DMR-1339564.

\bibliography{references}

\begin{thebibliography}{25}
\bibitem[{not()}]{note-orb}
\bibinfo{note}{Note that the pseudo-spin operator $\tau_z$, associated to the
  orbital degree of freedom, depends only on $n_1$ and $n_2$, and thus it does
  not couple to a magnetic field parallel to the QD.}
\bibitem[{gva()}]{gvalue}
\bibinfo{note}{The g-value used for this estimation was the experimental value
  $|g|=0.44$, while all plots in the paper were obtained for $g=2.0$.}
\end{thebibliography}

\setcounter{figure}{0}
\makeatletter 
\renewcommand{\thefigure}{S\@arabic\c@figure}

\section{Supplemental Material}

{\it Spin polarization: NRG results.}\textemdash 
In Fig.~\ref{figure1s}(a), NRG conductance results are shown along the (white) 
dashed line shown in Fig.~1(a) of the main text. The results are shown per channel, 
per spin. Channel 1, spin up and channel 2, spin down: open (red) circle; 
channel 2, spin up and channel 1, spin down: open (blue) square.
The peaks occur at the points $\alpha$ and $\beta$ described above. It is easy to see that the current polarization 
is reversed, between both channels, from points $\alpha$ to $\beta$. To see that 
more clearly, panel (b) shows the polarization, defined as $P_{\lambda}= 
(G_{\lambda \uparrow}-G_{\lambda \downarrow})/(G_{\lambda \uparrow}+G_{\lambda \downarrow})$, 
where $\lambda=1,2$ indicates the channels, and $\uparrow$, $\downarrow$ the spin orientation 
along the $z$-axis (parallel to the QD's 2d electron gas). The open (red) circles and 
(blue) squares are results for channels 1 and 2, respectively. These calculations were done 
for $U^{\prime}=0.1$, $\Gamma=0.027$, and $H=0.013$. 

\begin{figure}
  \begin{center}
    \includegraphics[width=3.0in]{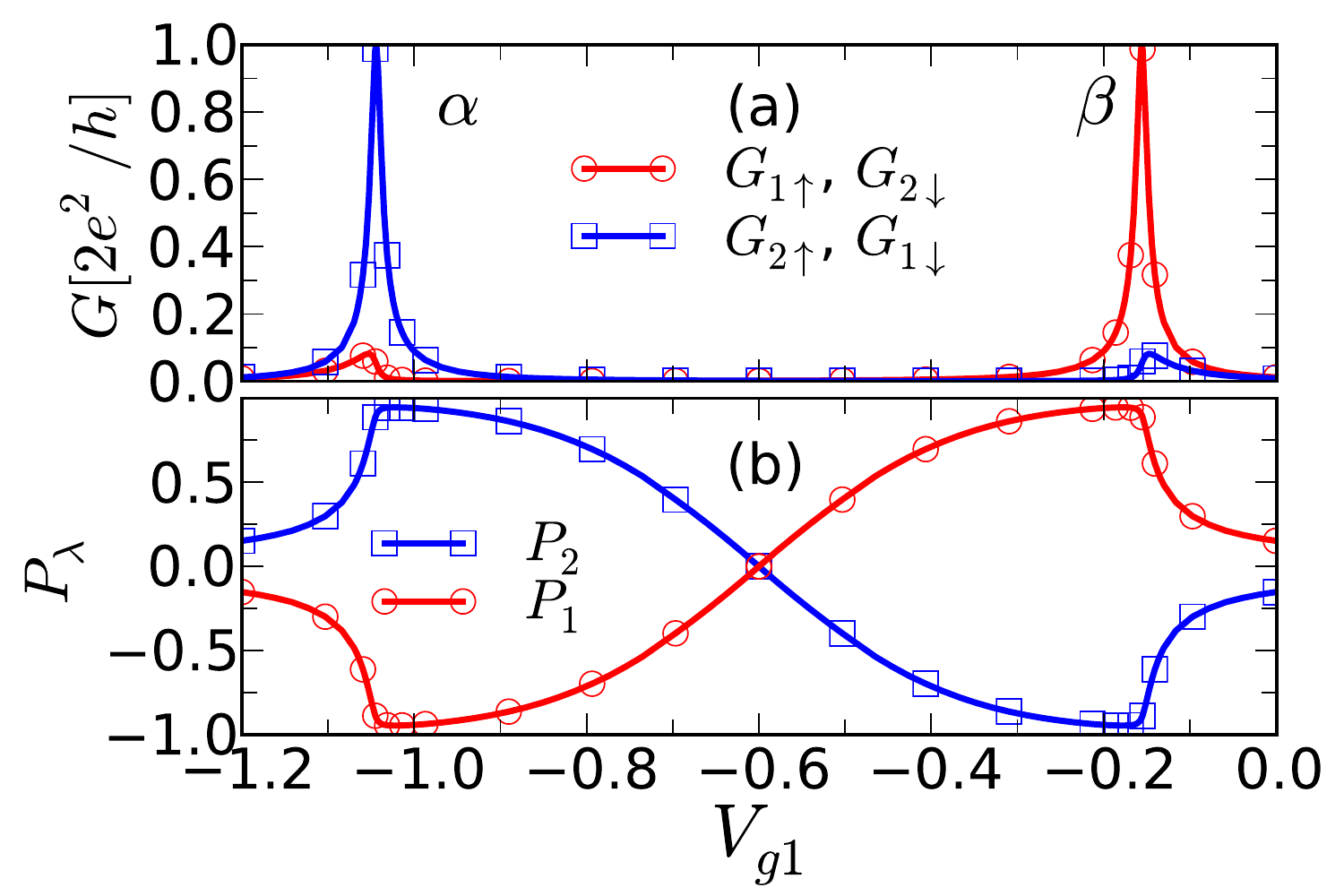}
  \end{center}
\caption{(Color online) (a) Conductance as a function 
of gate potential $V_{g1}$ for different channels 
and different spin orientations. Channel 1, spin up; channel 2, spin down: open (red) circle; 
channel 2, spin up; channel 1, spin down: open (blue) square. Results for $U^{\prime}=0.1$, $H=0.013$, 
and $\Gamma=0.027$. (b) Conductance polarization (see text for definition) for data 
in panel (a). Open (red) circles, channel 1; open (blue) squares, channel 2. 
}
\label{figure1s}
\end{figure}

\begin{figure}
  \begin{center}
   \vspace{.5cm}
    \includegraphics[width=3.25in]{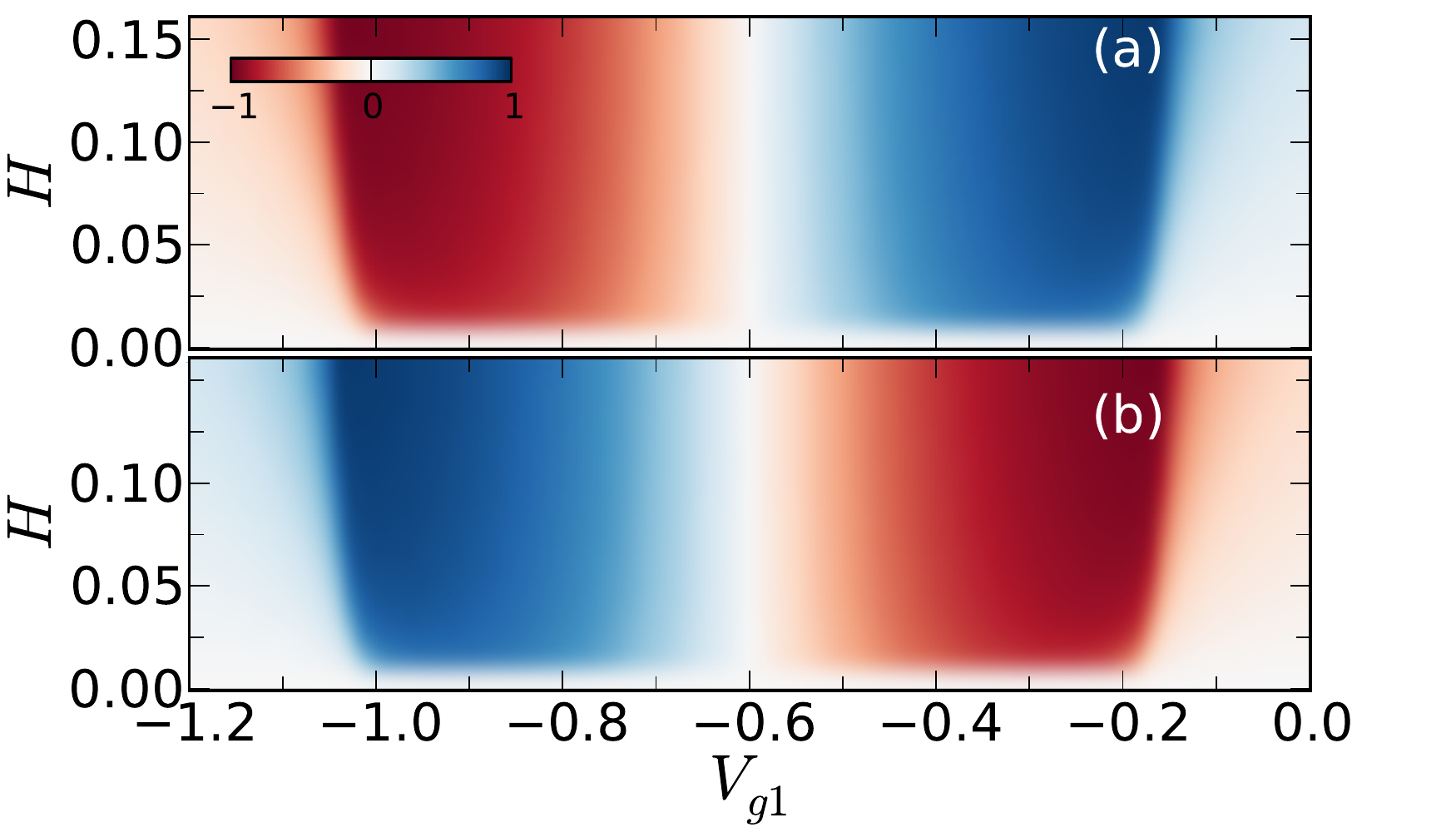}
  \end{center}
\caption{(Color online) Contour plot of the polarization 
as a function of the magnetic field $H$ (vertical axis) and the gate potential $V_{g1}$ (horizontal axis) 
[$V_{g1}$ is related to $V_{g2}$ by $V_{g2}=-V_{g1}-(U+2U^{\prime})$, i.e. the (white) dashed line 
in Fig.~1(a)]. Panels (a) and (b) correspond to channels 1 and 2, respectively.
}
\label{figure2s}
\end{figure}

{\it Spin polarization field dependence.}\textemdash 
To map out the polarization for 
different field values, Fig.~\ref{figure2s} shows a contour plot of $P_{\lambda}$ for 
$0.0 \leq H \leq 0.013$ for the same $V_{g1}-V_{g2}$ interval as in Fig.~\ref{figure1s} 
(panels (a) and (b) for channels 1 and 2, respectively). 
As already implied by the curves in Fig.~\ref{figure1s}(b), the high polarization region 
seems broad enough to be experimentally observable. As in the previously obtained DMRG 
results, a small value of field is enough to generate sizable polarization; the difference now is that 
a large polarization is obtained for a much wider interval along the dashed (white) line 
in Fig.~1(a). One should note however, by inspecting Fig.~\ref{figure1s}(a), that 
the $V_{g1}$ interval for which the conductance is close to unitary is narrow. As 
discussed in the main text, a large polarization can be sustained along a transversal direction in the 
$V_{g1}-V_{g2}$ plane [along the charge degeneracy line (yellow) shown in more detail in Fig.~1(b)].

\end{document}